\documentclass[aps,prl,floatfix,showpacs,twocolumn,tightenlines,superscriptaddress]{revtex4}

\usepackage{amsmath}
\usepackage{graphicx}
\usepackage{epsfig}

\begin{document}

\title{Entanglement generation in persistent current qubits}

\author{J.\ F.\ Ralph}\email{jfralph@liv.ac.uk}
\affiliation{Department of Electrical Engineering and Electronics, The University of Liverpool, 
Brownlow Hill, Liverpool, L69 3GJ, United Kingdom}

\author{T.\ D.\ Clark}
\affiliation{School of Engineering, The University of Sussex, 
Falmer, Brighton, BN1 9QT, United Kingdom. }

\author{T.\ P.\ Spiller}
\affiliation{Hewlett-Packard Laboratories, Filton Road, Stoke Gifford,
 Bristol BS34 8QZ, United Kingdom}

\author{W.\ J.\ Munro}
\affiliation{Hewlett-Packard Laboratories, Filton Road, Stoke Gifford,
Bristol BS34 8QZ, United Kingdom}

\date{September 2, 2004}

\begin{abstract}
In this paper we investigate the generation of entanglement between two persistent
current qubits. The qubits are coupled inductively to each other and to a common
bias field, which is used to control the qubit behaviour and is 
represented schematically by a linear oscillator mode. We consider
the use of classical and quantum representations for the qubit control fields 
and how fluctuations in the control fields tend to suppress entanglement. In particular,
we demonstrate how fluctuations in the bias fields affect the entanglement generated
between persistent current qubits and may limit the ability to design practical systems.
\end{abstract}


\maketitle

\section{Introduction}

\noindent
In this paper, we investigate the use of inductive coupling to generate entanglement
between two persistent current qubits. We are particularly interested in the 
representation of the magnetic bias fields that are used to control the behaviour of the qubits, 
and the requirements placed on these fields by the need
to generate significant levels of entanglement. Given the recent experimental results indicating coherent
quantum behaviour in superconducting persistent current and other Josephson devices 
\cite{Orl99,Fri00,Chi03}, the extension of these systems to arrays of coupled
qubits for quantum information processing is important and timely. Indeed, experiments
have already been reported using coupled persistent current qubits \cite{Ber03}. We
demonstrate that the requirements placed on the bias fields could present significant
obstacles to the use of persistant current qubits in quantum information processing.

We consider two models for the bias field: one classical 
and one quantum mechanical. In each model, the field is represented 
by a lossy linear oscillator, whose resonant frequency and coupling to 
the qubits can be varied. When the natural frequency of the field mode is significantly 
lower than the qubit frequencies, the conventional approach is to treat the bias
as a classical variable \cite{Orl99,Fri00,Chi03,Ber03} and to use the 
expectation value of the screening current in the classical equations of motion
when the bias dynamics are important \cite{Ral92,Spi92}. However, where
the bias field has fluctuations at frequencies that are comparable with the
qubit fluctuations, the classical model is no longer valid and the quantum
mechanical model predicts some interesting dynamical behaviour.
In addition, we derive a set of constraints for the accuracy of the bias fields 
which must be obeyed for a significant amount of entanglement to be produced. 
These constraints may limit the entanglement that can be produced in a practical
system. In particular, we derive restrictions on the coupling between the qubits and the
bias fields and the operating frequencies of multiple persistent current qubits.

\section{Coupled Persistent Current Qubits}

The persistent current qubits studied in this paper have been
proposed by Orlando et al. \cite{Orl99}. A schematic circuit
diagram is given in Figure 1, with inductive coupling between
the two qubits and the common control field.
\begin{figure}
\begin{center}
\includegraphics[height=7cm]{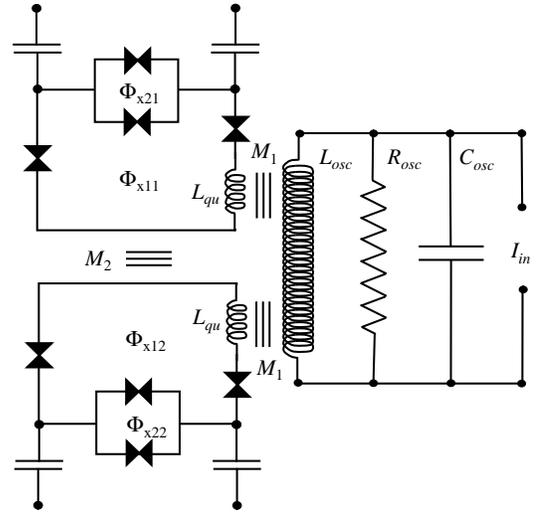}
\end{center}
\caption{Schematic diagram of coupled qubit system.}
\end{figure}  
The qubit inductance is negligible when compared with the effective inductance
generated by the series Josephson junctions in the loop. 
This means that the behaviour of an isolated qubit will tend to be dominated by the series 
Josephson junctions rather than the geometrical inductance of the ring, $L_{qu}\simeq 10$ pH. 
This allows the circuit to be simplified to a two-state model, 
corresponding to current states differing by approximately 600 nA \cite{Orl99}.
The two-state Hamiltonian for a single qubit, with two control fields 
$\Phi_{x1}$ and $\Phi_{x2}$, is given by \cite{Orl99},
\begin{equation}
\hat{H}_{qu}(\Phi_{x1},\Phi_{x2})=\left(\begin{array}{cc}
F(\Phi_{x1},\Phi_{x2})& -B(\Phi_{x1},\Phi_{x2}) \\
-B(\Phi_{x1},\Phi_{x2})& -F(\Phi_{x1},\Phi_{x2})
\end{array}\right)
\end{equation}
where the basis states, $\{|0\rangle, |1\rangle\}$, are the persistent current states with approximately $\pm 300$ nA, 
$\Phi_{x1}$ is the primary bias field for the main ring circuit and $\Phi_{x2}$ is a secondary
bias field that is used to modulate the critical current of the effective Josephson junction formed
by the two parallel junctions in the smaller secondary ring. The matrix elements are given by,
\begin{equation}
F(\Phi_{x1},\Phi_{x2})=r_1\left(\frac{\Phi_{x1}}{\Phi_0}\right)+r_2\left(\frac{\Phi_{x2}}{\Phi_0}\right)
\end{equation}
\begin{equation}
B(\Phi_{x1},\Phi_{x2})=\frac{t_1+s_1\left(\frac{\Phi_{x1}}{\Phi_0}\right)}{1-\eta
\sqrt{\frac{E_J}{E_C}}\left(\frac{\Phi_{x2}}{\Phi_0}\right)}
\end{equation}
and $\Phi_0=h/2e=2\times 10^{-15}$ Wb.
The circuit constants are taken from \cite{Orl99}: $r_1 = 2\pi E_J\sqrt{1-\frac{1}{4\beta^2}}=2r_2$, 
$s_1=0$, $t_1=0.001 E_J$, $\eta=3.5$, $\beta=0.8$,
$E_J\equiv 200$ GHz, $E_C=E_J/80$.

Although the energy level separation, and hence the dynamics of a single persistent current qubit, is dominated
by the Josephson energy of the junctions in the circuit, the inductance is important when determining the coupling 
between the qubit and the external fields and between the qubits themselves. For the system shown in Figure 1, 
this gives
\begin{eqnarray}\label{fluxcurrent}
\left(\begin{array}{c}
\Phi_{qu_1} \\
\Phi_{osc} \\
\Phi_{qu_2}
\end{array}
\right)
&=&\left(\begin{array}{ccc}
L_{qu}   & M_1       & M_2  \\
M_1      & L_{osc}   & M_1  \\
M_2      & M_1       & L_{qu}
\end{array}
\right)
\left(\begin{array}{c}
I_{qu_1} \\
I_{osc} \\
I_{qu_2}
\end{array}
\right) \nonumber\\
&=&{\bf M}\cdot
\left(\begin{array}{c}
I_{qu_1} \\
I_{osc} \\
I_{qu_2}
\end{array}
\right) \end{eqnarray}
where $M_1$ is the mutual inductance between the qubits and the bias coil and $M_2$ is the 
mutual inductance between the two qubits, $\Phi_{osc}$ is the magnetic flux in the shared bias field 
(which is treated as a linear oscillator and characterized by a capacitance $C_{osc}$ and an
inductance $L_{osc}$). In the absence of dissipation, the effective Hamiltonian for the
combined system can be written in the form \cite{Ral92},
\begin{eqnarray}
H&=&\frac{Q_{osc}^2}{2C_{osc}}+\frac{\Phi_{osc}^2}{2L_{osc}}-\Phi_{osc}I_{in} \nonumber\\
&&+\hat{H}_{qu_{1},qu_{2}}(\mu_1\Phi_{osc},\Phi_{x21};\mu_1\Phi_{osc},\Phi_{x22})
\end{eqnarray}
where $I_{in}$ is an external current used to fix the
static bias point (the oscillator fluctuates about this point), and
the coupling coefficients are given by $K_1^2=M_1^2/L_{qu}L_{osc}$,
$\mu_1=M_1/L_{osc}$ and $K_2^2=M_2^2/L_{qu}^2=\mu_2^2$. Each qubit has two bias/control fields, 
$\Phi_{x11}$ and $\Phi_{x21}$ for qubit 1 and $\Phi_{x12}$ and $\Phi_{x22}$ for qubit 2.
The primary control fields for the qubits, $\Phi_{x11}$ and $\Phi_{x12}$, are common
so we put $\Phi_{x11}=\Phi_{x21}=\mu_1\Phi_{osc}$.

To derive the Hamiltonian for the two qubits, we examine the energy of the
inductive circuit components. The Hamiltonian terms corresponding to the inductive
energies will have the form,
\begin{eqnarray}
H_{induc}&=&\frac{1}{2}(\Phi_{qu_1}\hspace{2mm} \Phi_{osc} \hspace{2mm} \Phi_{qu_2})
\cdot{\bf M}^{-1}\cdot
\left(\begin{array}{c}
\Phi_{qu_1} \\
\Phi_{osc} \\
\Phi_{qu_2}\end{array}\right) \nonumber \\
&=& \frac{1}{2}(I_{qu_1}\hspace{2mm} I_{osc} \hspace{2mm} I_{qu_2})
\cdot{\bf M}^T\cdot
\left(\begin{array}{c}
I_{qu_1} \\
I_{osc} \\
I_{qu_2}\end{array}\right)
\end{eqnarray}
Expanding the second of these expressions, the cross-coupling terms between the two
qubits has the form: $\Delta H_{qu_{1},qu_{2}} = \mu_2 L_{qu} \hat{I}_{qu_1}\hat{I}_{qu_2}$. 
The other terms corresponding to a shift in the effective self inductance of the qubits
and cross-coupling between the qubits and the oscillator are subsumed into the $F$ and 
$B$ terms. The two qubit Hamiltonian then has the form,
\begin{equation}\begin{array}{l}
\hat{H}_{qu_{1},qu_{2}}(\Phi_{x11},\Phi_{x21};\Phi_{x12},\Phi_{x22})=\\ \\
\left(\begin{tiny}\begin{array}{cccc}
F_1+F_2+\Delta_{12} & -B_2 & -B_1 & 0 \\
-B_2 & F_1-F_2-\Delta_{12} & 0 & -B_1 \\
-B_1 & 0 & -F_1+F_2-\Delta_{12} & -B_2 \\
0 & -B_1 & -B_2 & -F_1-F_2+\Delta_{12}   
\end{array}\end{tiny}\right)
\end{array}\end{equation}
in the current basis $\{|0_1 0_2\rangle, |0_1 1_2\rangle, |1_1 0_2\rangle, |1_1 1_2\rangle\}$
(which is used as the computational basis for the purposes of this paper), and where
$F_1=F(\Phi_{x11},\Phi_{x21})$, $B_1=B(\Phi_{x11},\Phi_{x21})$,
$F_2=F(\Phi_{x12},\Phi_{x22})$, $B_2=B(\Phi_{x12},\Phi_{x22})$.  
The $\Delta$ term comes from the qubit-qubit coupling term given in equation (6), with
$\Delta_{12} = K_2 L_{qu}\bar{I}_{qu}^2$, where $\bar{I}_{qu}\simeq 300$ nA 
is the magnitude of the screening current in the qubit logic (persistent current) states.

We assume that both qubits are identical and we consider the dynamics of a 
(common) primary control field $\Phi_{osc}$, keeping the secondary fields 
$\Phi_{x21}$ and $\Phi_{x22}$ fixed. Both the primary and secondary 
fields are nominally set to zero so that the energy eigenstates of the individual 
qubits are symmetric/anti-symmetric superpositions of 
the qubit current states. Initialising the
qubits in a current state will produce coherent oscillations at frequencies around
400 MHz. Although the fields are nominally zero, they all include a  
fixed error and the primary field includes the dynamics of the bias circuit.
(We use a common primary bias for computational simplicity and 
because the decoherence rate will be lower where any noise due to the flux bias 
is the same for each qubit). 

For simplicity, we assume that the couplings are  
weak, typically $K_1=0.002$ and $K_2=0.01$. This means that first order coupling terms will
be sufficient for most purposes. Allowing stronger couplings between the qubits and the bias fields
could introduce a range of problems: difficulties initialising the qubits
in a given state since the flux and current states are no longer identical
(due cross-couplings in equation (1)), and the quantum fluctuations in the
bias coil can affect the ability to generate entanglement (see below).

\section{Bias Fields and Dynamics}

We consider two models: one where the control field is a noisy classical
field and one where it is represented by a quantum oscillator. The classical oscillator model  \cite{Ral92,Spi92}
is expected to be valid as long as (i) the typical frequency of the oscillator is significantly 
lower than that of qubits, (ii) the (possibly entangled) two-qubit state
and the oscillator state are separable, and (iii) as long as the quantum fluctuations 
of an equivalent quantum oscillator (approximately given by the width of the energy 
eigenstates in a magnetic flux basis) are small compared to the other 
fluctuations that couple to the qubits. 
The quantum model uses a standard harmonic oscillator basis for the control field, 
and couples via an oscillator flux operator (formed from the raising and lowering operators, 
$\hat{a}^{\dagger}$ and $\hat{a}$) in the $F$ and $B$ functions. 

The classical approximation is based on the Born-Oppenheimer approximation that is often used
in nuclear and molecular physics. This removes the dynamics of a `fast' degree of freedom by replacing
the quantum mechanical operators with their expectation values; thereby averaging or integrating
out the effect of their dynamics on the the other `slow' degrees of freedom. The details of the
approximation and the restrictions on its use are more fully described in reference \cite{Spi92}.
In this case, the qubit behaviour is assumed to be fast compared to the evolution of the classical oscillator, 
and the expectation value of the energy is included in the (now classical) Hamiltonian given in equation (5).
The classical equation of motion is then derived in the conventional way using the variational derivative
with respect to the oscillator magnetic flux $\Phi_{osc}$, and the energy expectation value becomes
the expectation value of the combined qubit screening current. Using this approximation, and adding a 
parallel resistance $R_{osc}$, the equation of motion is given by \cite{Ral92,Spi92}, 
\begin{equation}\begin{array}{l}
C_{osc} \frac{d^2\Phi_{osc}}{dt^2}+\frac{1}{R_{osc}} \frac{d\Phi_{osc}}{dt}+\frac{\Phi_{osc}}{L_{osc}}
\\
=I_{in}+\mu_1\left<\hat{I}_{qu_1}(\mu_1\Phi_{osc},\Phi_{x21})+\hat{I}_{qu_2}(\mu_1\Phi_{osc},\Phi_{x22})\right>
\end{array}\end{equation}
where the qubit screening currents are calculated from the expectation value of the qubit screening 
current operators $\hat{I}_{qu_1}$ and $\hat{I}_{qu_2}$ over 
the instantaneous wavefunction (i.e. a pure state) of the two-qubit state (calculated
using the time-dependent Schr\"{o}dinger equation). The time-dependent Schr\"{o}dinger equation is used
for the qubit evolution in this case because, for simplicity, we assume that the dominant source of decohence 
is the oscillator and any intrinsic dissipation due to emission from the qubits in the cavity is comparatively 
small. However, the effect of this emssion process on the behaviour of a classical oscillator has been
examined elsewhere \cite{Ral03}. The dissipative term acts as a source of classical 
fluctuations due to Johnson noise in the resistor at finite temperature, taken to be $T=10$mK which is in line
with experimental systems. (The noise need not be thermal, but it is a useful generic model for
experimental noise because electronic noise is often characterised in
terms of an effective noise `temperature').
 
\section{Quantum Evolution and Quantum Jumps}
 
In the quantum model, the reduced density operator for the qubits is estimated
using a quantum trajectory model: an unravelling of the Markovian Master equation that 
produces individual `trajectories', which can then be averaged over an ensemble 
to produce an estimate of the density operator \cite{Car93,Gis93,Wis96}. (A recent comprehensive 
review of this subject is given in reference \cite{Ple98}). Each 
unravelling is equivalent to the Master equation when averaged over an ensemble, but corresponds 
to a different measurement interaction at the individual system level \cite{Wis96}.  
For simplicity, we choose the `quantum jumps' model \cite{Car93} and 
thermal environment (Lindblad) operators for the oscillator described in \cite{Sar74}, 
\begin{equation}
\hat{L}_1=[(\bar{n}+1)\omega_{osc}Q_{osc}]^{\frac{1}{2}}\hat{a} \hspace{0.5cm}
\hat{L}_2=[\bar{n}\omega_{osc}Q_{osc}]^{\frac{1}{2}}\hat{a}^{\dagger}
\end{equation}
where $\bar{n}=[\exp(\hbar\omega_{osc}/kT)-1]^{-1}$ is the thermal oscillator
occupancy, $\omega_{osc}=1/\sqrt{C_{osc}L_{osc}}$ is the resonant frequency of the bias feld 
and the quality factor is given by $Q_{osc}=\omega_{osc}R_{osc}C_{osc}$ 
(in this paper we use $Q_{osc}=200$ for both the classical and quantum models). These environmental
operators represent emission and absorption of photons from the environment by the oscillator mode, which
is assumed to be the dominant source of dissipation in this paper.

The quantum jump evolution is calculated by numerically integrating the full state (describing the
qubits and the oscillator) over discrete time intervals (of size $dt$) and applying three different 
evolution operators. In each time interval, there
is a small (but finite) probability that the bias oscillator will emit or absorb a quantum of energy
from the environment. The probabilities for emission and absorption during the time step are found from,
\begin{equation}
P_1(dt) = \left\langle \hat{L}_1^\dagger \hat{L}_1\right\rangle dt = [(\bar{n}+1)\omega_{osc}Q_{osc}]
\left\langle \hat{a}^\dagger \hat{a}\right\rangle dt
\end{equation}
\begin{equation}
P_2(dt) = \left\langle \hat{L}_2^\dagger \hat{L}_2\right\rangle dt = [\bar{n}\omega_{osc}Q_{osc}]
\left\langle \hat{a} \hat{a}^{\dagger}\right\rangle dt
\end{equation}
The jumps are generated stochastically and when a jump occurs, a projection
operator is applied to the instantaneous state of the system. If an emission occurs,
an operator $\hat{\Omega}_1(dt)=\sqrt{dt}\hat{L}_1$ is applied, lowering the state of the oscillator, 
and if absorption occurs an operator $\hat{\Omega}_2(dt)=\sqrt{dt}\hat{L}_2$ is applied, raising the 
oscillator state. The state is then renormalised. In the absence of a quantum jump, the 
evolution of the system is found from the non-unitary evolution operator,
\begin{equation}
\hat{\Omega}_0(dt) = 1-\frac{i dt}{\hbar}\hat{H} - \frac{dt}{2}(\hat{L}_1^\dagger \hat{L}_1
+ \hat{L}_2^\dagger \hat{L}_2)
\end{equation}
The non-unitary term is added to ensure that the evolution of the density operator for the coupled
system agrees with that predicted by the Markovian Master equation, when the (pure state) density operators
generated from an ensemble of individual `trajectories' are averaged to produce an estimate for the 
mixed state density matrix for the whole system $\rho_{total}$: two qubits and bias oscillator. The reduced density
operator for the two qubits $\rho_{1+2}$ is then found by performing a partial trace over the oscillator states. 
(The validity of the quantum jumps approach can be checked by selecting a different unravelling for
the Master equation, and in this paper the results have been verified by comparing the behaviour
obtained from the quantum jumps unravelling to equivalent results obtained from the quantum
state diffusion unravelling \cite{Gis93}. Quantum state diffusion produces a continuous stochastic
evolution of the quantum state, and can be shown to correspond to a unit-efficiency heterodyne
detection measurement process \cite{Wis96}).

\section{Results}

In each model, quantum and classical, the initial conditions set for the oscillator are a thermalised state:
a thermal quantum state for the quantum oscillator and an initial condition generated from the classical
equation of motion, which has been allowed to come into equilibrium with the thermal noise by numerically
integrating its behaviour prior to initialisation using a different realisation of the noise for each qubit
trajectory calculation. The qubits are initialised in a product of pure current states.
The initial states of the qubits are chosen to be either in-phase
($|0_1 0_2\rangle$) or anti-phase ($|0_1 1_2\rangle$). That is, the 
coherent oscillations induced by initialising each qubit in a current
state are initially in phase with each other or in anti-phase. Each individual
quantum trajectory is calculated using the same initial conditions for the qubit states with
different static bias errors for the control fields. The static bias errors are fixed for
each trajectory and represent the accuracy with which the fields might be set in
an experiment. The size of these static bias errors is found to be crucial to the generation 
of useable entangement between the qubits. The entanglement is characterised in terms
of the {\it concurrence}, which is widely used in quantum information processing for bipartite
systems \cite{Woo97}. The concurrence for the two-qubit mixed state density matrix $\rho_{1+2}$ is defined as
$$
C(\rho_{1+2})=max\{0,\sqrt{\lambda_1}-\sqrt{\lambda_2}-\sqrt{\lambda_3}-\sqrt{\lambda_4}\}
$$
where $\sqrt{\lambda_1},\ldots,\sqrt{\lambda_4}$ are the eigenvalues of
the matrix $\rho_{1+2}(\sigma_y\otimes\sigma_y)\rho_{1+2}^*(\sigma_y\otimes\sigma_y)$
in nondecreasing order and $\sigma_y$ is a Pauli spin matrix \cite{Wei03}.
Although the concurrence is used here, other measures of entanglement can be
calculated from this: e.g. the entanglement of formation $E_F$ can be calculated from
$$
E_F(\rho_{1+2})=h\left(\frac{1}{2}[1+\sqrt{1-C(\rho_{1+2})^2}]\right)
$$
where
$$
h(x) = -x\log_2 x-(1-x)\log_2 (1-x)
$$
at least for this two-qubit system.
The {\it mixedness} of the resultant two-qubit state is given by the von Neumann entropy of the 
reduced density operator \cite{Wei03}: $S(\rho_{1+2}) = Tr(\rho_{1+2} \log_4 \rho_{1+2})$. The logarithm 
is taken to base 4 because the qubit states exist in a four-dimensional Hilbert space, giving a 
mixedness parameter that varies between zero and one.

Figure 2 shows the concurrence of the two-qubit mixed states 
for bias errors of $\Delta\Phi_{x}=10^{-5}\Phi_0$ (1 $\sigma$, Gaussian)
for both models for an oscillator frequency
of 100 MHz. We see that the concurrence oscillates and decays gradually in time, mainly due to
dephasing between the two qubits originating from the static error in the bias fields, and approaches
a maximially entangled mixed state in each oscillation. 
By varying the size of the errors, we find that the concurrence is quite sensitive 
to errors in the static fields, bias errors larger than about  
$5\times 10^{-5}\Phi_0$ do not lead to any useful entanglement in the qubits - the concurrence 
is less than 0.3 at all times and decays very rapidly. 
(Other calculations away from the minimum splitting point of the qubits, where
the frequency differences between energy eigenstates are higher, indicate that the requirements
on the bias fields are even more demanding. Where the qubit frequencies are around 1-2 GHz,
useful entanglement is only generated if the bias errors are less than about $\Delta\Phi_{x}\sim 10^{-6}\Phi_0$).
\begin{figure}
\begin{center}
\includegraphics[height=12cm]{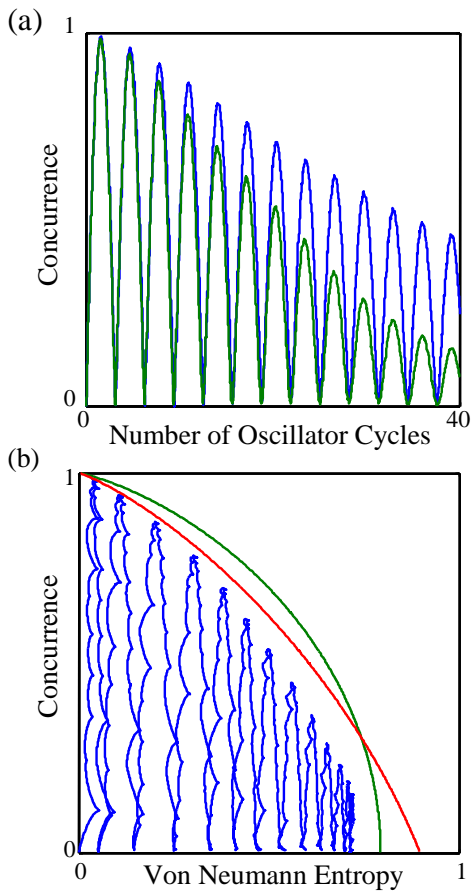}
\end{center}
\caption{(a) Concurrence versus number of oscillator
cycles (100MHz) for both the classical (blue)
and quantum (green) field models, (b) Concurrence versus von Neumann 
entropy for mixed state (blue) corresponding to the
quantum model shown in (a), with two types of maximally
entangled mixed states (Rank 3 (green) and Werner states (red) \cite{Wei03}).}
\end{figure}

By increasing the couplings between the two qubits, the rate at which the two qubits
become entangled can be increased. However, this could lead
to problems when initialising the qubit states since the qubit current and flux
are mixed by the inductive coupling. Care is required to ensure that the initialisation
process projects the states onto the correct basis.
(Slight differences in the initial states will affect the entanglement
in the mixed state, but this is not explicitly considered here).
Increasing the coupling between the oscillator and the qubits is likely to lead to additional
problems. Although the classical oscillator model contains noise due to thermal 
fluctuations and dissipation from the finite quality factor of the oscillator, a 
quantum oscillator also includes quantum fluctuations. The differences
in Figure 2(a) for the quantum and classical oscillator models are due to the comparatively
low Q value used and the quantum fluctuations coupling across to the qubits. Increasing the 
oscillator Q and/or reducing the oscillator frequency improves the agreement between
classical and quantum models. The coupling between the bias and the qubits is sufficiently 
small for the entanglement between the oscillator and qubits to be negligible. 

This raises an interesting point: what happens when the quantum fluctuations in the 
oscillator coupling across to the qubits are comparable with the static bias errors? 
The size of the quantum fluctuations in the oscillator can be estimated from the flux 
width of the harmonic oscillator states. 
Using the width of the oscillator states and the coupling
coefficient, the approximate size of the fluctuations that couple to the qubits will be,
\begin{equation}
\mu_1\Delta\Phi_{osc} \simeq K_1\sqrt{2\hbar\omega_{osc}L_{qu}} \sim 1.2\times 10^{-6}\Phi_0
\end{equation}
for $K_1=0.002$ and an oscillator frequency of 100 MHz. The size of the 
quantum fluctuations that couple across varies linearly with $K_1$ and
as the square root of the frequency. This means that the field frequencies 
must be very low if strong field-qubit couplings are to be used.
Keeping the frequency constant and increasing the oscillator-qubit coupling we find that
as soon as the quantum fluctuations become comparable with the constraints on the
bias errors, entanglement is effectively lost ($\sim 10^{-5}\Phi_0$ for the cases 
considered here and $\sim 10^{-6}\Phi_0$ for qubits biased away from the minimum splitting 
point). Even in the absence of the static bias errors, the quantum noise will affect
the generation of entangement between the two qubits. Figure 3 shows the effect of
increasing the size of the quantum noise by increasing the coupling to the bias field.
For a 100 MHz oscillator and a coupling of $K_1=0.01$, giving fluctuations
$\mu_1\Delta\Phi_{osc} \simeq 6\times 10^{-6}\Phi_0$, the entanglement between the
two qubits is lost very quickly. That this is due to quantum fluctuations, rather than
the change in coupling strength alone, can be verified by simultaneously changing the
oscillator frequency and the coupling strength keeping the size of the quantum
noise given by equation (13) fixed.
\begin{figure}
\begin{center}
\includegraphics[height=6cm]{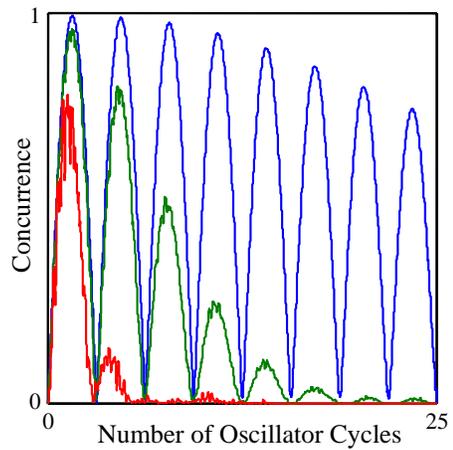}
\end{center}
\caption{Concurrence versus number of oscillator
cycles (100MHz) for quantum field model with different coupling strengths
in the absence of static bias errors: 
$K_1=0.002$ (blue), $K_1=0.005$ (green), $K_1=0.01$ (red).}
\end{figure}

The quantum fluctuations effectively limit the operating frequency of persistent current qubits as 
quantum processing devices, because the operating frequency must be lower than the frequency 
at which the bias fields may be manipulated, which is determined by the frequency and
the quality factor of the bias circuit. Increasing the operating frequency of the device,
and keeping the fluctuations below the required level, would mean reducing the coupling between the
qubits and the applied field, which might make it difficult to address individual elements 
of an array of qubits. 

\begin{figure}
\begin{center}
\includegraphics[height=9cm]{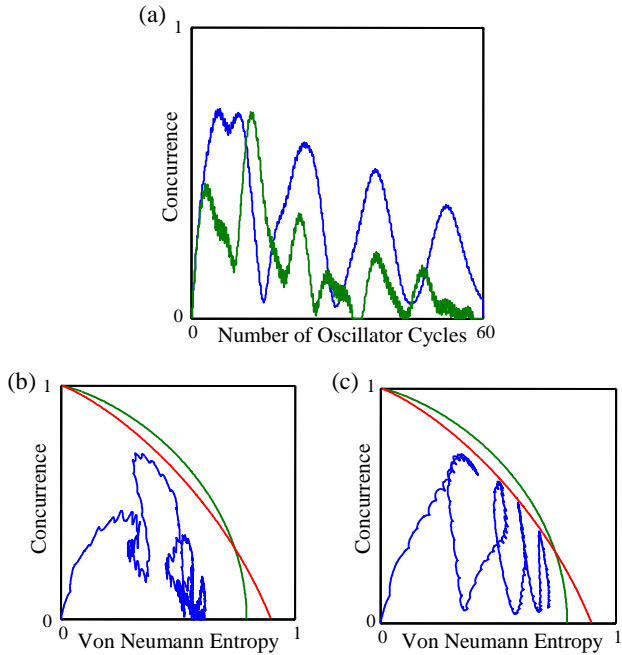}
\end{center}
\caption{(a) Concurrence versus number of oscillator
cycles (380 MHz) using the quantum model for qubits
initialised in the $|0_1 1_2\rangle$ state (blue) and the $|0_1 0_2\rangle$ state (green); 
(b) Concurrence versus von Neumann entropy for qubits initialised in the $|0_1 0_2\rangle$ state (blue), 
with two types of maximally entangled mixed states (Rank 3 (green) and Werner states (red) 
\cite{Wei03}); (c) as (b) for the $|0_1 1_2\rangle$ state.}
\end{figure}
In spite of the possible difficulties in biasing and addressing individual qubits within an array, 
there are some aspects of the behaviour of this tripartite system that are worth investigating further.
In particular, in situations where the frequencies of the oscillator and the qubits are not widely
separated, it is possible to generate interesting evolution whereby the currents flowing in the
qubits excite oscillations in the bias field, which modifies the behaviour of the qubits. Figure
4 shows examples for a quantum oscillator with a natural frequency of 380 MHz, for both the in-phase 
and anti-phase initial states (all other parameters are identical to Figure 
2). For the $|0_1 0_2\rangle$ initial state, the screening currents flowing in the qubits add in-phase.
The net current coupled to the oscillator acts as a sinusoidal drive which excites
oscillations in the bias field. In the $|0_1 1_2\rangle$ initial state the net current coupling to the oscillator 
is close to zero initially and the concurrence oscillations are far more regular. There are significant differences
in the concurrence and the von Neumann entanglement between the two cases. In particular, the entanglement
persists for longer for the $|0_1 1_2\rangle$ initial state and the von Neumann entropy exhibits some large scale
oscillations, as shown by the `loops' in Figures 4(b) and 4(c). These oscillations in entropy 
correspond to points where the oscillations in the bias field are at their largest. 
The oscillations in the field shift the bias point of both of the qubits, which accentuates the natural 
dephasing between the qubits. As the relative phase of the qubit oscillations changes, the net current coupling
to the oscillator changes, and - as they approach the anti-phase state - the net current coupling to the
oscillator falls and the oscillations in the field reduce, thereby stabilising the relative phase
of the coherent oscillations. Although these phase slips occur for the anti-phase initial
state, they are more evident for the in-phase initial state and are responsible for the rapid decay
of entanglement in this case and are the dominant source of decoherence.

\section{Conclusions}

In this paper, we have discussed a coupled system consisting of two persistent
current qubits and a linear oscillator, representing one of the qubit control fields. 
We have examined the generation of entanglement between 
the two qubits in the presence of a dynamical bias field, and have shown that
classical models are approximately valid for low frequency fields
with high quality factors. We have used the models to set constraints on the accuracy of the 
applied control fields. The static bias errors must be less than about $5\times 10^{-5}\Phi_0$ for
the parameters used in this paper.
We have also considered cases where the underlying quantum fluctuations 
in the applied field are significant, and have used this to derive a constraint that
relates the coupling between the bias field and the qubits and the frequencies present in the bias fields. 
If these constraints are not met, the useful entanglement between the two qubits is effectively lost.
This could affect the use of these devices in a practical quantum processing system, placing
severe demands on the accuracy of the static control fields and limiting the operating frequencies
of these devices.
However, we have found that, where the frequencies of the applied fields and the qubits are
comparable, some interesting dynamical behaviour can be produced by the back 
reaction of the qubits on the applied field.



\begin{thebibliography}{999}

\bibitem{Orl99}
T.P.Orlando, J.E.Mooji, L.Tian, C.H. van der Wal, L.S.Levitov,
S.Lloyd, J.J.Mazo, {\it Phys. Rev. B} {\bf 60}, 15398 (1999).
\bibitem{Fri00}
J.R.Friedman, V.Patel, W.Chen, S.K.Tolpygo, J.E.Lukens, 
{\it Nature} {\bf 406}, 43 (2000);
C.H. van der Wal, A.C.J. ter Haar, F.K.Wilhem, R.N.Schouten,
C.J.P.M.Harmans, T.P.Orlando, S.Lloyd, J.E.Mooij, {\it Science} 
{\bf 290}, 773 (2000);
J.M.Martinis et al. {\it Phys. Rev. Lett.} {\bf 89}, 117901 (2002)
\bibitem{Chi03}
I.Chiorescu, Y.Nakamura, C.J.P.M.Harmans, J.E.Mooij, {\it Science}
{\bf 299}, 1869 (2003).
\bibitem{Ber03}
A.J.Berkley,H.Xu, R.C.Ramos, M.A.Gubrud, F.W.Strauch, P.R.Johnson,
J.R.Anderson, A.J.Dragt, C.J.Lobb, F.C.Wellstood, {\it Science} {\bf 300}, 1548 (2003);
A.Izmalkov, M.Grajcar, E.Il'ichev, Th. Wagner, H.-G.Meyer,
A.Yu.Smirnov, M.H.S.Amin, Alec Maassen van den Brink, A.M.Zagoskin, cond-mat/0312332v1, December 2003.
\bibitem{Ral92}
J.F.Ralph, T.P.Spiller, T.D.Clark, R.J.Prance, H.Prance, {\it Int. J. Mod. Phys. B} {\bf 8}, 2637 (1994);
J.Diggins, J.F.Ralph, T.D.Clark, T.P.Spiller, R.J.Prance, H.Prance, {\it Phys. Rev. E} {\bf 49}, 1854 (1994);
T.D.Clark, J.F.Ralph, R.J.Prance, H.Prance, J.Diggins, R.Whiteman, {\it Phys. Rev. E} {\bf 57}, 4035 (1998).
\bibitem{Spi92}
T.P.Spiller, T.D.Clark, R.J.Prance, H.Prance, {\it Phys. Lett. A} {\bf 170}, 273 (1992).
\bibitem{Ral03}
J.F.Ralph, T.D.Clark, M.J.Everitt, H.Prance, P.Stiffell, R.J.Prance,
{\it Phys. Lett. A} {\bf 317}, 199, (2003).
\bibitem{Car93}
H.J.Carmichael, `An Open System Approach to Quantum Optics' (Lecture
Notes in Physics, Vol.18), Springer-Verlag, Berlin, 1993.
\bibitem{Gis93}
N.Gisin, I.C.Percival, {\it J. Phys. A} {\bf 26}, 2233 (1993);
N.Gisin, I.C.Percival, {\it J. Phys. A} {\bf 26}, 2246 (1993);
G.C.Hegerfeldt, {\it Phys. Rev. A} {\bf 47}, 449 (1993).
\bibitem{Wis96}
H.M.Wiseman, {\it Quant. Semiclass. Opt.} {\bf 8}, 205 (1996).
\bibitem{Ple98}
M.B.Plenio, P.L.Knight, {\it Rev. Mod. Phys.} {\bf 70}, 101 (1998).
\bibitem{Sar74}
M.Sargent III, M.O.Scully, W.E.Lamb Jr., `Laser Physics', Ch.16, Addison-Wesley, 1974.
\bibitem{Woo97}
W.K.Wooters, {\it Phys. Rev. Lett.} {\bf 80}, 2245 (1997).
\bibitem{Wei03}
T.-C.Wei, K.Nemoto, P.M.Goldbart, P.G.Kwiat, W.J.Munro, F.Verstraete,
{\it Phys. Rev. A} {\bf 67}, 022110 (2003).

\end{thebibliography}
\end{document}